\documentclass[prd,twocolumn,showpacs,nofootinbib,amsfonts,amssymb,amsmath]{revtex4}

\usepackage{graphicx}
\usepackage{simplewick}
\usepackage{epstopdf}
\usepackage{multirow}
\usepackage{cases}
\usepackage{bm}
\usepackage{epsfig}
\usepackage{color}

\def\be{\begin{equation}}
\def\ee{\end{equation}}
\def\bea{\begin{eqnarray}}
\def\eea{\end{eqnarray}}

\newcommand{\bear}{\begin{eqnarray}}

\newcommand{\eear}{\end{eqnarray}}

\newlength{\tskip}
\newbox\pippobox

\def\be{\begin{equation}}
\def\ee{\end{equation}}
\def\bea{\begin{eqnarray}}
\def\eea{\end{eqnarray}}

\def\9{\nabla}

\def\dd{{\rm d}}

\def\6{\partial}

\def\0{(0)}

\def\>{\rightarrow}

\newcommand{\PP}{{\cal P}}

\begin{document}

\title{{\bf Reconstruction of the primordial power spectra with Planck and} BICEP2}

\author{Bin Hu$^{1}$, Jian-Wei Hu$^{2}$, Zong-Kuan Guo$^{2}$, Rong-Gen Cai$^{2}$}
\email{hu@lorentz.leidenuniv.nl}
\email{jwhu@itp.ac.cn}
\email{guozk@itp.ac.cn}
\email{cairg@itp.ac.cn}
\affiliation{
$~^{1}$Institute Lorentz of Theoretical Physics, University of Leiden, 2333CA Leiden, The Netherlands\\
$~^{2}$State Key Laboratory of Theoretical Physics, Institute of
Theoretical Physics, Chinese Academy of Sciences, P.O. Box 2735, Beijing 100190, China}

\begin{abstract}
By using the cubic spline interpolation method,
we reconstruct the shape of the primordial scalar and tensor power spectra
from the recently released {\it Planck} temperature and BICEP2 polarization
cosmic microwave background data. We find that
the vanishing scalar index running ($\dd n_s/\dd\ln k$) model is strongly disfavored
at more than $3\sigma$ confidence level on the $k=0.0002$ Mpc$^{-1}$ scale.
Furthermore, the power-law parameterization gives a blue-tilt tensor spectrum, no matter
using only the first 5 bandpowers $n_t = 1.20^{+0.56}_{-0.64}~(95\% {\rm CL})$
or the full 9 bandpowers $n_t = 1.24^{+0.51}_{-0.58}~(95\% {\rm CL})$ of BICEP2 data sets.
Unlike the large tensor-to-scalar ratio value ($r\sim0.20$)
under the scale-invariant tensor spectrum assumption, our interpolation approach gives
$r_{0.002} < 0.060~(95\% {\rm CL})$ by using the first 5 bandpowers of BICEP2 data.
After comparing the results with/without BICEP2 data, we find that
{\it Planck} temperature with small tensor amplitude signals and BICEP2 polarization data
with large tensor amplitude signals dominate the tensor spectrum reconstruction on 
the large and small scales, respectively. 
Hence, the resulting blue tensor tilt actually 
reflects the tension between {\it Planck} and BICEP2 data.
\end{abstract}


\maketitle


\section{Introduction}
Recently BICEP2 experiment \cite{Ade:2014xna} reported an excess of Cosmic Microwave Background (CMB)
B-mode polarization power spectrum over the base lensed-$\Lambda$CDM expectation in the range $30<\ell<150$, inconsistent
with the null hypothesis at a significance of $>5\sigma$. Since the single field slow-roll inflationary model
predicts a peak around multipole $\ell\sim80$ in the B-mode
auto-correlation (BB) spectrum seeded by the primordial gravitational wave/tensor perturbation mode,
the BICEP2 results are believed as the first indirect detection of the primordial gravitational wave.
Under the assumption of  power-law scalar and scale-invariant tensor spectra,
the observed B-mode power spectrum is well-described by a lensed-$\Lambda$CDM+tensor theoretical model with tensor-to-scalar ratio
$r=0.20^{+0.07}_{-0.05}$ with $r=0$ disfavored at $7.0\sigma$ confidence level.

However, the scientific results of BICEP2 data are in tension with those from other CMB experiments,
such as {\it Planck} \cite{Ade:2013uln}.
The first discrepancy is in the amplitude of the scale-invariant tensor spectrum, which is described by
the tensor-to-scalar ratio $r\equiv A_t/A_s$. Unlike the scalar perturbatoins,
due to the absence of acoustic oscillation mechanism the tensor contributions to the temperature CMB spectrum are rapidly washed out
inside the horizon at electron-proton recombination epoch ($\ell\ge 200$) \cite{Hu:1997hp}. Hence, the temperature anisotropies
on the large scales are the mixture of scalar and tensor contributions. Furthermore, if one assumes the simple power-law
form of the primordial scalar power spectrum, i.e., no scalar index running $\dd n_s/\dd \ln k=0$,
the precisely measured higher multipoles by {\it Planck} put stringent constraints on the scalar amplitude $A_s$ and index $n_s$.
Therefore, in order to explain the observed power deficit in the low-$\ell$ regime by {\it Planck}, one has to suppress the
tensor spectrum amplitude. Consequently, from only the temperature anisotropy measured by {\it Planck}, one has
$r<0.11$ at $95\%$, which is in a ``very significant'' tension (around $0.1\%$ unlikely) with BICEP2 results \cite{Smith:2014kka}.
As stressed by the BICEP2 team, however, this tension could be reconciled by adding the running of the scalar index which
is the degree of freedom to suppress the scalar temperature anisotropy in low-$\ell$ regime. Then, combining {\it Planck}
with WMAP low-$\ell$ polarization \cite{Hinshaw:2012aka}
and ACT \cite{Das:2013zf} /SPT \cite{Keisler:2011aw,Story:2012wx,Reichardt:2011yv} high-$\ell$ data, one could get $r<0.26$ at $95\%$.
Besides that, several other possible solutions to this tension have been proposed,
such as step feature spectra \cite{Bousso:2014jca,Miranda:2014wga,Hazra:2014aea},
fast-slow roll model \cite{Hazra:2014jka}, anti-correlation
scalar iso-curvature initial condition \cite{Kawasaki:2014lqa,Kawasaki:2014fwa},
sterile neutrino species \cite{Giusarma:2014zza,Zhang:2014dxk,Dvorkin:2014lea},
sudden change in speed of inflaton or Lorentz violation \cite{Contaldi:2014zua}. And see also
\cite{Freivogel:2014hca,Firouzjahi:2014fda,Cai:2014bda,Cheng:2014bta,Choudhury:2014kma,Hotchkiss:2011gz} for other possibilities. 

The second tension is between the observed blue-tilt tensor spectrum ($n_t>0$)
\cite{Gerbino:2014eqa,Wu:2014qxa,Cheng:2014ota,Li:2014cka,Chang:2014loa}
by BICEP2 and the red-tilt one ($n_t=-r/8$) predicted by the standard inflationary paradigm.
Generally, the blue tensor spectrum asks for violation of Null Energy Condition (NEC),
which is equivalent to $\rho+P<0$ (or $\dot H>0$) in the flat universe.
There exist several NEC violation inflationary models in the literature, such as
super-inflation \cite{Baldi:2005gk}, phantom inflation \cite{Piao:2004tq}, G-inflation \cite{Kobayashi:2010cm} {\it etc}.
Some alternative paradigm of inflation, such as ``string gas cosmology'' \cite{Brandenberger:2014faa},
bouncing universe
\cite{Creminelli:2006xe,Creminelli:2007aq,Buchbinder:2007ad,Xia:2014tda,Qiu:2014nla,Piao:2003zm,Liu:2013kea} or
other possibilities \cite{Gong:2014qga,Wang:2014kqa} (see the references therein)
might be helpful to solve this tension.

In this paper, we start from a purely phenomenological point of view to reconstruct the shape of primordial scalar and tensor
spectra from {\it Planck} temperature and BICEP2 polarization data.

\section{Parameterization of primordial spectra and Datasets}
In order to reconstruct a smooth spectrum with the continuous
first and second derivatives, in this paper we adopt the cubic spline interpolation method,
which has been used to analyze WMAP or {\it Planck} temperature and polarization data
\cite{Sealfon:2005em,Verde:2008zza,Peiris:2009wp,Guo:2011re,Guo:2011hy,Guo:2012jn,Aslanyan:2014mqa,Abazajian:2014tqa}.
Beside the cubic spline interpolation reconstruction, there also exist other model independent algorithms, such as
Bayesian evidence selected linear interpolation, \cite{Vazquez:2012ux,Vazquez:2013dva} {\it etc.}
Due to the fact that the amplitude of CMB anisotropy is so tiny, $\delta T/\bar T\sim \mathcal O(10^{-5})$,
and CMB observational windows cover several orders of magnitude in
spatial scale, it is reasonable to parameterize the logarithms of primordial spectra, which seed the
CMB anisotropy, in the logarithms of fluctuation wavenumber, $\ln k$.
The method of cubic spline interpolation can be summarized as follows.

First, we uniformly sample $N_{\rm bin}$ points in the logarithmic scale of wavenumber.
Second, inside of the sampled bins $\ln k_i < \ln k < \ln k_{i+1}$, we use the
cubic spline interpolation to determine logarithmic values of the primordial
power spectrum.
Third, the boundary conditions are adopted, where the second
derivative is set to zero.
For $k<k_1$ or $k>k_{N_{\rm bin}}$ we fix the slope of
the primordial power spectrum at the boundaries and linearly extrapolate to
the outside regimes. Mathematically, the corresponding formula could be written
as
\begin{widetext}
\bea
\label{pps_form}
\ln \PP(k) = \left\{
\begin{array}{ll}
\left.\frac{\dd\ln \PP(k)}{\dd\ln k}\right|_{k_1} \ln \frac{k}{k_1} + \ln \PP(k_1), &  k<k_1; \\
\ln \PP(k_i), & k\in \{k_i\}; \\
{\rm cubic \ spline}, & k_i < k < k_{i+1};\\
\left.\frac{\dd\ln \PP(k)}{\dd\ln k}\right|_{k_{N_{\rm bin}}} \ln \frac{k}{k_{N_{\rm bin}}} +
\ln \PP(k_{N_{\rm bin}}), & k>k_{N_{\rm bin}}.
\end{array}
\right.
\eea
\end{widetext}
This reconstruction method has three advantages:
first of all, it is easy to detect deviations from a scale-invariant or
a power-law spectrum because both the scale-invariant and
power-law spectra are just straight lines
in the $\ln k$-$\ln \PP$ plane.
Second, negative values of the spectrum can be avoided by using $\ln \PP(k)$
instead of $\PP(k)$ for splines with steep slopes.
Finally, the shape of the power spectrum reduces to the scale-invariant or power-law
spectrum as a special case when $N_{\rm bin}=1,2$, respectively.

Since the purpose of this work is to reconstruct the scalar and tensor spectra,
we need to adopt different sampling logarithms based on the different observational windows.
For the primordial scalar curvature spectrum, its constraints are mainly driven by the CMB
temperature modes. With {\it Planck} sensitivities, we uniformly sample 3 bins ranged in
$\ln k\in(-8.517,-1.609)$, which corresponds to $k \in (0.0002,0.2)$ Mpc$^{-1}$.
For the tensor spectrum, we adopt two uniformly logarithmic sampling strategies, one is corresponding
to the scales $k\in(0.002,0.03)$ Mpc$^{-1}$, the other is $k\in(0.002,0.02)$ Mpc$^{-1}$.
This is because the BICEP2 B-mode polarization data, which is a very sensitive probe for the
primordial tensor spectrum, have an excess of B-mode power in all the range of polar angle mulitpoles ($20\leq\ell\leq340$).
And moreover, compared with the first 5 bandpowers, which is in the range of ($20\leq\ell\leq200$), the power in
the second 4 bandpowers has extraordinary excess over the base lensed-$\Lambda$CDM expectation.
This extraordinary excess might arise from exotic physical origin beyond standard inflationary paradigm or from
some unresolved foreground contaminations. Given this consideration, in this work we take two different choices
of BICEP2 data, the first is using the full 9 bandpowers, and the second is to use the selected first 5 bandpowers.
Hence, we have to adjust our sampling logarithms as mentioned above.

In the left part of this section, we would like to briefly review the data sets we used.
First of all, we utilize the {\it Planck} TT power spectra, namely, for low-$\ell$ modes ($2\leq\ell<50$)
via all the 9 frequency channels ranged from $30\sim353$ GHz, for high-$\ell$ modes ($50\leq\ell\leq2500$)
through $100$, $143$, and $217$ GHz frequency channels \footnote{\url{http://pla.esac.esa.int/pla/aio/planckProducts.html}} \cite{Ade:2013zuv}.
Second, in order to break the well-known parameter degeneracy between the re-ionization optical depth and the amplitude of
CMB temperature anisotropy, we also include WMAP9 low-$\ell$ temperature/polarization spectra ($2\leq\ell\leq32$) \cite{Hinshaw:2012aka}.
In addition, we use BICEP2 polarization (EE,EB,BB) spectra
from 9 (or 5) bandpowers of multipoles in ($20\leq\ell\leq340~{\rm or}~200$) of 150 GHz channels \cite{Ade:2014xna}
\footnote{\url{http://www.cfa.harvard.edu/CMB/bicep2/papers.html}}.
For the data analysis numerical package,
we compute the CMB angular power spectra by using
the public Einstein-Boltzmann solver \texttt{CAMB} \cite{Lewis:1999bs}
and explore the cosmological parameter space with a Markov Chain Monte Carlo sampler,
namely \texttt{CosmoMC} \cite{Lewis:2002ah}.

\section{Results and Discussions}
In this section we will start with the
scalar spectrum reconstruction, and then turn to the
tensor spectrum case.
The prior ranges of the primordial spectrum parameters 
we studied are listed in Tab.\ref{tab:parameters}. Here we 
emphasize that the differences in the prior of tensor spectra amplitude at our cubic 
spline sampling knots when BICEP2 data are included, 
lie in the tension between {\it Planck} and BICEP2 data (we will show later). 
When our MCMC sampler investigates the wide parameter space spanned by ($\ln B_1,\ln B_2,\ln B_3$), at 
some points the resulting spectra are inconsistent with {\it Planck} TE cross-correlation
data, in order to avoid this problem we have to adjust the prior ranges.
But still the width of the priors are large enough and also the tensor amplitudes 
($\ln B_1,\ln B_2,\ln B_3$) get well constrained in these prior ranges as shown in 
Fig.\ref{Fig:tri_tensor_mnew}. So, we conclude that our prior choices will not affect the results significantly.
\begin{table}
\begin{tabular}{c|c}
\hline\hline
Parameter & Range (min, max)  \\
\hline
$\ln (10^{10} A_s^2)$ &  $(2.7, 4.0)$ \\
$n_s$ &  $(0.9, 1.1)$ \\
$\dd n_s/\dd \ln k$ & $(-1.0, 1.0)$ \\
$r_{0.05}$ & $(0.0, 2.0)$ \\
$n_t$ & $(-1.0, 5.0)$ \\
\hline
$\ln (10^{10} A_1)$ & $(1.0, 5.0)$ \\
$\ln (10^{10} A_2)$ & $(1.0, 5.0)$ \\
$\ln (10^{10} A_3)$ & $(1.0, 5.0)$ \\
$\ln (10^{10} B_1)^{\ast}$ & $(-3.0, 3.0)~/(-3.0, 3.0)$ \\
$\ln (10^{10} B_2)^{\ast}$ & $(-2.0, 5.0)~/(-3.0, 3.0)$ \\
$\ln (10^{10} B_3)^{\ast}$ & $(-1.0, 5.0)~/(-3.0, 3.0)$ \\
\hline\hline
\end{tabular}
\caption{List of the primordial spectrum parameters used in the Monte Carlo sampling.
~${^{\ast}}$The left parameter ranges are for the chains from {\it Planck}+WP+BICEP2 data compilation, 
and the right one are for those without BICEP2 data.}
 \label{tab:parameters}
\end{table}

\subsection{Scalar spectrum reconstruction}
Since the main capability of scalar spectrum reconstruction is driven by the
CMB temperature data, we firstly study the case without BICEP2 polarization, i.e.
only with {\it Planck} temperature and WMAP9 low-$\ell$ polarization (WP) data sets.
As mentioned in the previous section, here we uniformly sample 3 points
in the logarithmic scale of wavenumber, which are located at $k_1=0.0002$, $k_2=0.0063$
and $k_3=0.2$ Mpc$^{-1}$ with the logarithmic amplitudes $\ln A_1$, $\ln A_2$ and $\ln A_3$, respectively.
And then, we sample the parameter space spanned by vanilla $\Lambda$CDM parameters without
the scalar amplitude $\ln A_s$ and its index tilt $n_s$ and replacing them with $\ln A_1$, $\ln A_2$ and $\ln A_3$,
hereafter we call this parameter compilation as $\Lambda$CDM-$\ln A_s$-$n_s$+$\ln A_1$+$\ln A_2$+$\ln A_3$.
\begin{figure}[ht]
\begin{center}
\includegraphics[width=0.4\textwidth]{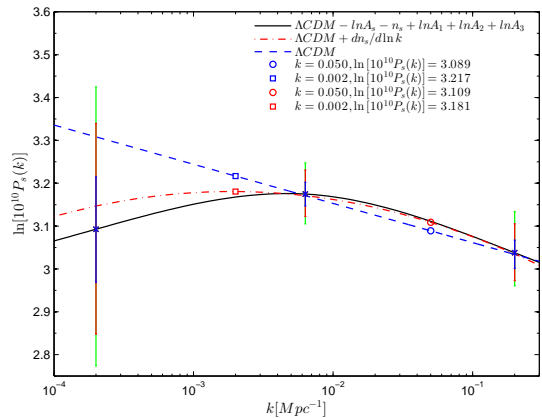}
\caption{Reconstruction of primordial scalar spectrum without BICEP2 data.}
\label{Fig:scalar_shape_noBICEP2}
\end{center}
\end{figure}

The marginalized mean scalar spectrum reconstructed from {\it Planck}+WP data is represented by the black solid curve
in Fig.~\ref{Fig:scalar_shape_noBICEP2}
and the corresponding $1\sim3\sigma$ error bars at the sampling points are denoted by the blue, red and green segments, respectively.
For comparison, we also show the primordial scalar spectrum from the {\it Planck} marginalized mean vanilla $\Lambda$CDM
and $\Lambda$CDM+$\dd n_s/\dd\ln k$ (scalar index running) with blue dashed and red dotted-dashed curves.
From Fig.~\ref{Fig:scalar_shape_noBICEP2}, we can see that, first,
our cubic spline interpolation result mimics the $\Lambda$CDM+$\dd n_s/\dd\ln k$ case;
second, on $k=0.0002$ Mpc$^{-1}$ scale, the simplest vanilla model is disfavored at nearly $2\sigma$ level.

\begin{figure}[ht]
\begin{center}
\includegraphics[width=0.4\textwidth]{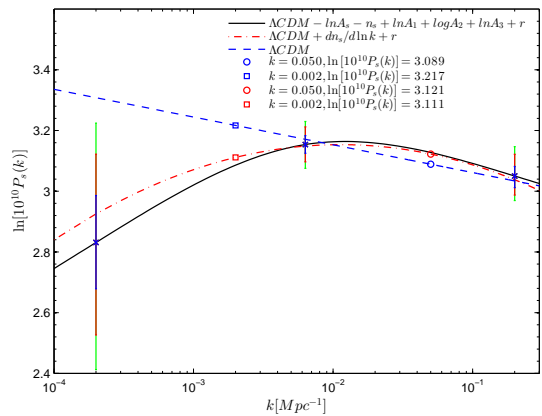}
\caption{Reconstruction of primordial scalar spectrum with BICEP2 data.}
\label{Fig:scalar_shape_BICEP2}
\end{center}
\end{figure}

Adding the BICEP2 polarization data, we show the results in Fig.~\ref{Fig:scalar_shape_BICEP2} for
models including the tensor-to-scalar ratio $r$, i.e.
$\Lambda$CDM-$\ln A_s$-$n_s$+$r$+$\ln A_1$+$\ln A_2$+$\ln A_3$.
We find that, first of all,
due to the anti-correlation between scalar and tensor amplitudes
(see the bottom left sub-panel of Fig.~\ref{Fig:tri_scalar} in Appendix~\ref{App:scalar}),
the large value of tensor-to-scalar ratio
discovered by BICEP2 data will lead to the suppression of scalar amplitude on the large scales. This is also
explicitly demonstrated in the top sub-panel of Fig.~\ref{Fig:tri_scalar} in Appendix~\ref{App:scalar}. As a result of deficit of scalar
power on the large scales, as shown in Fig.~\ref{Fig:scalar_shape_BICEP2} the vanilla $\Lambda$CDM model (blue curve) is strongly
disfavored with $>3\sigma$ confidence level on the $k=0.0002$ Mpc$^{-1}$ scale.
A similar result is obtained by authors of \cite{Abazajian:2014tqa}.
By using {\it Planck} and BICEP2 data, they
found a distinct preference for a suppression of power in the scalar spectrum at large
scales, $k\leq 10^{-3}$ Mpc$^{-1}$ via a linear spline reconstruction method.
Second, by assuming the scale invariant tensor spectrum, our scalar spectrum cubic spline interpolation parameterization
still gives a large tensor-to-scalar ratio $r=0.21^{+0.10}_{-0.09}$ at $95\%$C.L., see Tab.~\ref{Tab:AAA} in Appendix~\ref{App:scalar}.

\subsection{Tensor spectrum reconstruction}

In the previous subsection, we have assumed the tensor spectrum is scale-invariant.
In this subsection we relax the assumption and use the same cubic spline interpolation method to
reconstruct the shape of tensor spectrum. Unlike the CMB temperature spectrum which is mainly sourced
by the primordial scalar perturbations, the B-mode polarization anisotropy seeded by tensor perturbation
is only detected by BICEP2 on the large scales with the polar spherical harmonic mulitpoles
ranged $20\leq\ell\leq340$ (9 bandpowers).
As the case for the scalar spectrum, here the tensor spectrum is also uniformly sampled with 3 points
in the logarithmic scale of wavenumber, but only on the scales covered by BICEP2 observations.
Consequently, we sample them at $k_1=0.002$, $k_2=0.0077$ and $k_3=0.03$ Mpc$^{-1}$, respectively. The cosmological
parameters we estimated are the 6 vanilla $\Lambda$CDM model parameters plus the extra 3 tensor amplitudes
$\ln B_1$, $\ln B_2$ and $\ln B_3$.

\begin{figure}[ht]
\begin{center}
\includegraphics[width=0.4\textwidth]{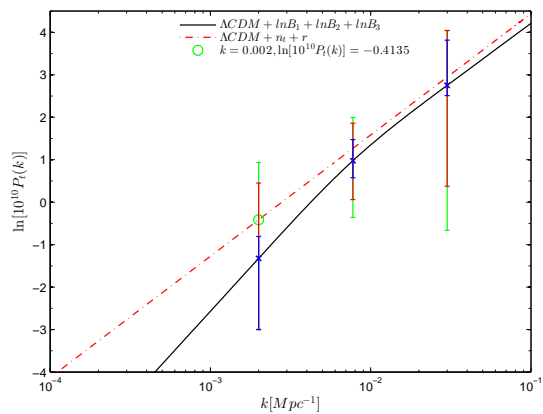}
\caption{Reconstruction of primordial tensor spectrum with BICEP2 data (9 bandpowers).}
\label{Fig:tensor_shape_9band}
\end{center}
\end{figure}

In Fig.~\ref{Fig:tensor_shape_9band} we plot the primordial tensor spectra from the best-fit
model of $\Lambda$CDM+$\ln B_1$+$\ln B_2$+$\ln B_3$ (black solid curve) and $\Lambda$CDM+$r$+$n_t$ (red dotted-dashed curve)
as well as the error bars at the sampling points of the cubic spline interpolation method.
First of all, both the standard power-law and our cubic spline parameterizations favor a blue-tilt
tensor spectrum. The former (see Tab.\ref{Tab:6rnt} in Appendix\ref{App:tensor}) reports
\bea
r_{0.002} &<& 0.061\;,~95\%{\rm CL}\;,~({\rm 9~bandpowers})\;,\label{r_9}\\
n_t &=& 1.24^{+0.51}_{-0.58}\;, 95\%{\rm CL}\;, ~({\rm 9~bandpowers})\;,\label{nt_9}
\eea
and our cubic spline interpolation method gives
\be
\label{pps_r02_9}
r_{0.002} < 0.064\;,95\%{\rm CL}\;,~({\rm 9~bandpowers})\;.
\ee
Second, we notice that in our cubic spline interpolation method the slope of the
tensor spectrum becomes larger in the low-$k$ regime, but still is consistent with
the power-law parameterization in $2\sigma$ confidence level.

As we argued in the previous section, there exists an extraordinary power excess
in the higher wavenumber regimes of BICEP2 data.
Given this consideration, in what follows we only adopt the selected first 5 bandpower
data of BICEP2 for our reconstruction. Based on the multipole ranges covered by  these
powerbands ($20\leq\ell\leq200$), we sample points at $k_1=0.002$, $k_2=0.0063$ and $k_3=0.02$ Mpc$^{-1}$, respectively.

\begin{figure}[ht]
\begin{center}
\includegraphics[width=0.4\textwidth]{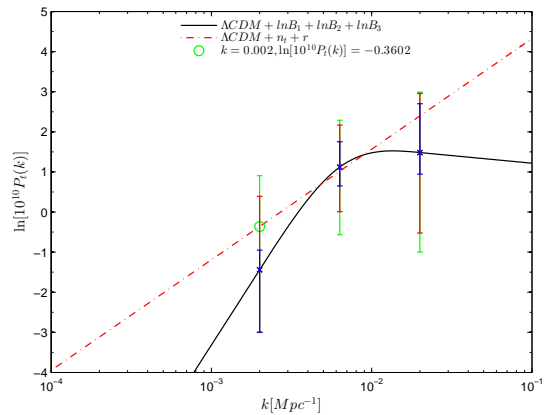}
\caption{Reconstruction of primordial tensor spectrum with BICEP2 data (5 bandpowers).}
\label{Fig:tensor_shape_5band}
\end{center}
\end{figure}

The reconstructed tensor spectra as well as error bars are shown in Fig.~\ref{Fig:tensor_shape_5band}.
First, from the power-law parameterization,
we can see that the blue tensor spectra are still favored but with the tilt becomes smaller as expected
\bea
r_{0.002} &<& 0.067\;,95\%{\rm CL}\;,~({\rm 5~bandpowers})\;,\label{r_5}\\
n_t &=& 1.20^{+0.56}_{-0.64}\;, 95\%{\rm CL}\;.~({\rm 5~bandpowers})\label{nt_5}
\eea
Second, with only the first 5 bandpower data, unlike the simplest power-law parameterization,
a non-trivial shape of tensor spectrum is obtained. Concretely, in the range of $k\in(0.002,0.006)$
BICEP2 data favor a large tensor blue-tilt, while when $k>0.0063$Mpc$^{-1}$ the spectrum becomes almost flat.
This is due to the fact that we do not use the last 4 bandpower data.
The resulting tensor-to-scalar ratio in our
cubic spline interpolation method is
\be
\label{pps_r02_5}
r_{0.002} < 0.060\;, 95\%{\rm CL}\;, ~({\rm 5~bandpowers})\;.
\ee

\begin{figure}[ht]
\begin{center}
\includegraphics[width=0.4\textwidth]{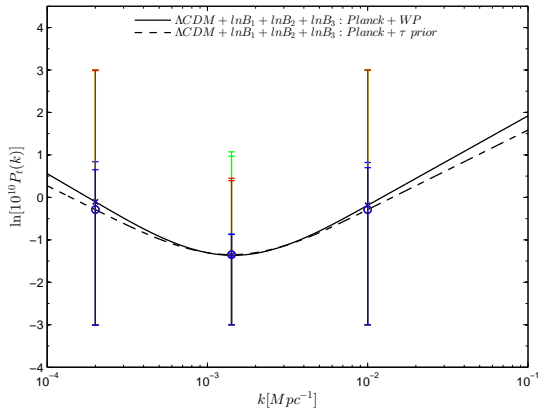}
\caption{Reconstruction of primordial tensor spectrum with {\it Planck}+WP and {\it Planck}+$\tau$ prior.}
\label{Fig:tensor_shape_NoBICEP2}
\end{center}
\end{figure}

Because the above blue tensor tilt is significantly inconsistent with the standard 
inflationary prediction, we have to figure out its reason on the data analysis level. Notice that 
for now we always utilize {\it Planck}+WP+BICEP2 data compilation, so one natural guess is that this blue 
tensor tilt might reflect the tension among {\it Planck}, WMAP polarization and BICEP2 data sets. 
In order to justify our conjecture, we have to remove the above data sets one by one. 
Since we vary both the primordial spectrum and standard $\Lambda$CDM parameters, 
such as baryon ($\Omega_b h^2$), cold dark matter density ($\Omega_c h^2$) {\it etc}, 
we have to keep the robust {\it Planck} temperature data in order to get well estimations of the 
$\Lambda$CDM parameters. Hence, we first remove the BICEP2, i.e. using {\it Planck}+WP, and further
discard WMAP polarization data, i.e. only using {\it Planck} temperature data. However, due to the 
fact that CMB temperature data are insensitive to the reionization optical depth ($\tau$), if we discard 
WMAP polarization data we have to include gaussian prior on $\tau$ to break the well-known degeneracy between 
$\tau$ and the scalar amplitude $A_s$. Here we take the gaussian prior as 
\be
\tau = 0.089\pm 0.013\;.
\ee

Besides this, because the contribution of tensor spectra to CMB temperature anisotropies is only significant
in the mulitpole range ($2\leq\ell\leq100$), when we use {\it Planck}+WP or {\it Planck}+$\tau$ prior data sets, 
we sample the $k$ knots of tensor spectra in the range of $(0.0002,0.01)$.

\begin{figure}[ht]
\begin{center}
\includegraphics[width=0.4\textwidth]{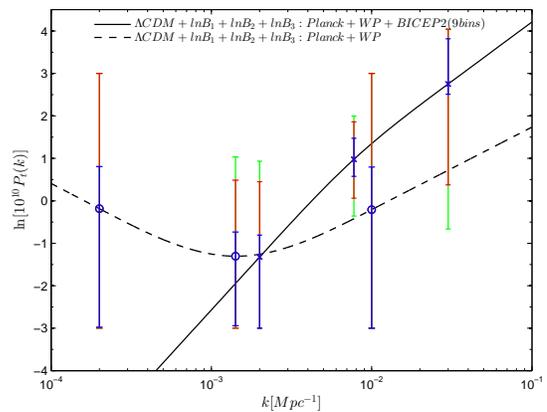}
\caption{Reconstruction of primordial tensor spectrum with/without BICEP2.}
\label{Fig:tensor_shape_wwout}
\end{center}
\end{figure}

The resulting primordial tensor spectrum shape, corresponding
marginalized 1D/2D posterior distributions and parameter regimes are shown in 
Fig.\ref{Fig:tensor_shape_NoBICEP2}, Fig.\ref{Fig:tri_tensor_nobicep2} and Tab.\ref{Tab:BBB_NoBICEP2}. 
First, we can see that the reconstructed tensor spectra from {\it Planck}+WP and {\it Planck}+$\tau$ prior 
are very similar, both the central values and the marginalized error bars.
So, we can draw the conclusion that WMAP low-$\ell$ polarization data are not crucial for the 
tensor reconstruction results. Second, as shown in Fig.\ref{Fig:tensor_shape_wwout} the error bars 
from {\it Planck} temperature data are quite large compared with those from the data compilation 
{\it Planck}+WP+BICEP2 (see the error bars at sampling knot $k=0.01$Mpc$^{-1}$ in the dashed black curve and
those at the second knot in the black solid curve in Fig.\ref{Fig:tensor_shape_wwout}). 
It means that the current {\it Planck} temperature data are not rubost to determine
the shape of the primordial tensor spectrum. The resulting tensor spectrum from only {\it Planck} temperature
data could be any shape among red, blue or scale-invariant types. 
Third, when we compare the second knot in {\it Planck}+WP (dashed black curve) 
and the first left knot {\it Planck}+WP+BICEP2 results (solid black curve),
Fig.\ref{Fig:tensor_shape_wwout} shows that both their central value and marginalized error bars are very close. 
It reflects the fact that the reconstructed tensor spectrum in the low-$k$ regime is actually driven by {\it Planck}
temperature data. Furthermore, considering the fact that the BICEP2 data dominate the high-$k$ part, 
we conclude that our reconstructed blue tensor tilt are due to the tension between {\it Planck} and BICEP2 data sets,
i.e. the fact that small tensor amplitude signals from {\it Planck} temperature data dominates the large scale reconstruction,
while the large tensor amplitude signals from BICEP2 B-mode polarization data dominates the small scale reconstruction, 
leads to the resulting blue tensor tilt.

\begin{figure}[ht]
\begin{center}
\includegraphics[width=0.4\textwidth]{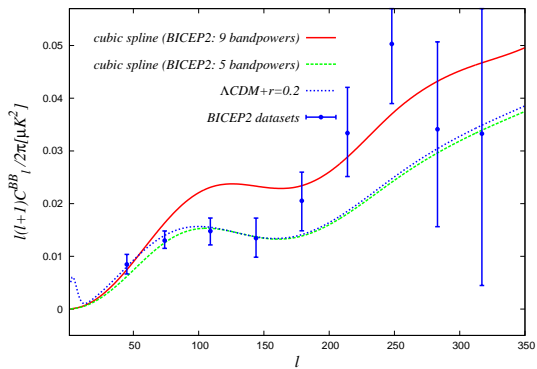}
\caption{CMB B-mode polarization auto-correlation spectrum $C_{\ell}^{BB}$.}
\label{Fig:clbb}
\end{center}
\end{figure}

Finally, in order to give an intuitive impression of our reconstruction result,
in Fig.~\ref{Fig:clbb}  we plot the BB auto-correlation power spectrum of our marginalized mean models
(listed in Tab.~\ref{Tab:BBB} of Appendix~\ref{App:tensor})
as well as the scale-invariant tensor spectrum with $r=0.2$ model against the BICEP2 bandpowers data sets.
We can see that in order to fit the BICEP2 data in the last 4 bandpowers,
compared with the 5 bandpower reconstruction result (green curve) and the scale-invariant one (blue),
the 9 bandpower (red) curve grows up in the high-$\ell$ regime significantly.

\section{Conclusions}
Starting with a purely phenomenological point of view, in this paper we
have reconstructed the shape of the primordial scalar and tensor spectra by using the cubic spline
interpolation method with {\it Planck} temperature and BICEP2 B-mode polarization data sets.
We find that, due to the anti-correlation between scalar and tensor amplitudes on the large scales,
the large value of tensor-to-scalar ratio
discovered by BICEP2 data will lead to the suppression of scalar amplitude in this regime.
Concretely, the vanishing scalar index running model is strongly disfavored by {\it Planck}+WP+BICEP2 data
compilation with more than $3\sigma$ confidence level on the $k=0.0002$ Mpc$^{-1}$ scale.
Furthermore, for the tensor spectrum reconstruction, a blue-tilt spectrum is obtained no matter
using only the first 5 bandpowers $n_t = 1.20^{+0.56}_{-0.64}~(95\%{\rm CL})$
or the full 9 bandpowers $n_t = 1.24^{+0.51}_{-0.58}~(95\%{\rm CL})$ of BICEP2 data sets.
Because of the large tensor tilt, compared with the large tensor-to-scalar ratio value ($r\sim0.20$)
under the scale-invariant assumption, our cubic spline interpolation method gives
$r_{0.002} < 0.060~(95\%{\rm CL})$ and $r_{0.002} < 0.064~(95\%{\rm CL})$
by using the data sets {\it Planck}+WP+BICEP2 (5 bandpowers) and (9 bandpowers), respectively.
Finally, we also studied the data without BICEP2, we found that our resulting 
blue tensor tilt actually reflects the tension in the tensor amplitude between {\it Planck} 
(small amplitude but dominate the reconstruction on the large scale) 
and BICEP2 (large amplitude but dominate the reconstruction on the small scale) data sets.

Our results show that the conclusion of the blue-tilt tensor spectrum is very significant and
independent of using power-law or cubic spline parameterizations. More important, this blue-tilt spectrum
is not consistent with the prediction of the standard single field inflationary paradigm $n_t=-r/8$. 
On the one hand, it asks for a more careful cross-check with future experiments,
such as the polarization data of {\it Planck} and Keck Array.
On the other hand, once this discovery is confirmed, it will lead to a paradigm revolution about our understanding
of the early universe.

\vspace{10cm}

\begin{acknowledgments}
BH are indebted to Ana Ach\'ucarro, Sabino Matarrese, Nicola Bartolo, Wessel Valkenburg
and Frederico Arroja for various helpful discussions.
BH is supported by the Dutch Foundation for Fundamental Research on Matter (FOM).
JWH, ZKG and RGC are partially supported by the project of Knowledge
Innovation Program of Chinese Academy of Science,
NSFC under Grant No.11175225, No.11335012, and National Basic Research
Program of China under Grant No.2010CB832805 and No.2010CB833004.
\end{acknowledgments}

\appendix
\section{Marginalized statistics in scalar spectrum reconstruction}
\label{App:scalar}
Here we list the various marginalized statistical results for
cubic spline interpolation and power-law parameterizations of scalar spectrum, including
1D, 2D marginalized posterior distribution, marginalized mean values
as well as the $68\%$ (or $95\%$) confidence levels.

\begin{figure*}[tmb]
\begin{center}
\includegraphics[scale=0.3]{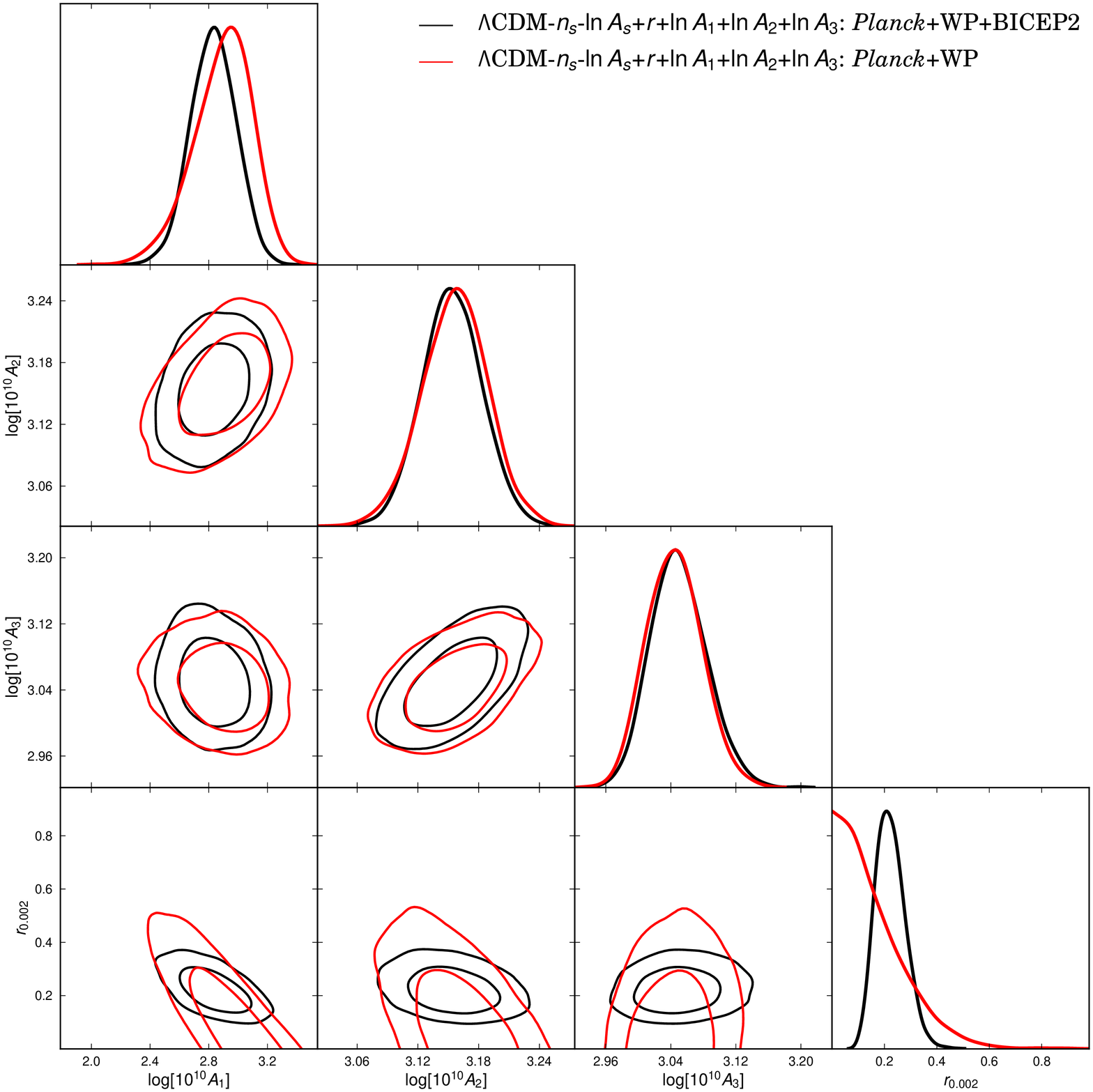}
\caption{1D/2D posterior distribution of scalar spectrum reconstruction.}
\label{Fig:tri_scalar}
\end{center}
\end{figure*}

\setlength\tabcolsep{1pt}
\begin{table*}[htb!]
\footnotesize
\centering
\begin{tabular}{|l|c|c|c|}
\hline
\multicolumn{3}{|c}{$\Lambda$CDM-$n_s$-$\ln A_s$+$r$+$\ln A_1$+$\ln A_2$+$\ln A_3$} &
\multicolumn{1}{|c|}{$\Lambda$CDM+$\dd n_s/\dd\ln k$+$r$}\\
\hline
\hline
                                            &  {\it Planck}+WP                  & {\it Planck}+WP+BICEP2 (9 bandpowers) & {\it Planck}+WP+BICEP2 (9 bandpowers)\\ 
\hline
Parameters                                  & mean $\pm$ $68\%$ C.L.            & mean $\pm$ $68\%$ C.L.  & mean $\pm$ $68\%$ C.L. \\ \hline
$100\Omega_b h^2$                           & 2.224$\pm$0.031                   & 2.224$\pm$0.030         & 2.238$\pm$0.028 \\
$\Omega_c h^2$                              & 0.1204$\pm$0.0028                 & 0.1202$\pm$0.0027       & 0.1186$\pm$0.0017 \\
$100\theta_{\rm MC}$                                 & 1.04125$\pm$0.00065               & 1.04133$\pm$0.00063     & 1.04150$\pm$0.00057 \\
$\tau$                                      & 0.103$\pm$0.016                   & 0.106$\pm$0.017         & 0.105$\pm$0.016 \\
$\ln(10^{10} A_s)$                         & $\cdots\cdots$                    & $\cdots\cdots$          & 3.122$\pm$0.033 \\
$n_s$                                       & $\cdots\cdots$                    & $\cdots\cdots$          & 0.9600$\pm$0.0063 \\
$\dd n_s/\dd\ln k$                       & $\cdots\cdots$                    & $\cdots\cdots$          & $-0.028^{+0.019}_{-0.021}$ ($95\%$CL) \\
$r$                                         & $<0.41$ ($95\%$CL)      & $0.21^{+0.10}_{-0.09}$ ($95\%$CL) & $0.20^{+0.08}_{-0.09}$ ($95\%$CL) \\
$\ln(10^{10} A_1)$                         & 2.89$\pm$0.20                     & 2.83$\pm$0.15           & $\cdots\cdots$ \\
$\ln(10^{10} A_2)$                         & 3.157$\pm$0.032                   & 3.154$\pm$0.029         & $\cdots\cdots$ \\
$\ln(10^{10} A_3)$                         & 3.045$\pm$0.034                   & 3.050$\pm$0.034         & $\cdots\cdots$ \\
\hline
$\Omega_{m}$                                & 0.318$\pm$0.018                   & 0.316$\pm$0.017         & 0.306$\pm$0.010 \\
$H_0 [\mathrm{km}/\mathrm{s}/\mathrm{Mpc}]$ & 67.22$\pm$1.27                    & 67.34$\pm$1.22          & 68.06$\pm$0.79 \\
\hline
\hline
$\chi^2_{\rm min}/2$                        & 4901.833                          & 4921.868                & 4924.052 \\
\hline
\end{tabular}
\caption{
Mean values and $68\%$ (or $95\%$) confidence limits for primary/derived parameters in the cubic spline 
and power-law parameterization of scalar spectrum.}
\label{Tab:AAA}
\end{table*}

\section{Marginalized statistics in tensor spectrum reconstruction}
\label{App:tensor}
Here we list the various marginalized statistical results for
cubic spline interpolation and power-law parameterizations of tensor spectrum, including
1D, 2D marginalized posterior distribution, marginalized mean values
as well as the $68\%$ (or $95\%$) confidence levels.
\begin{figure*}[tmb]
\begin{center}
\includegraphics[scale=0.25]{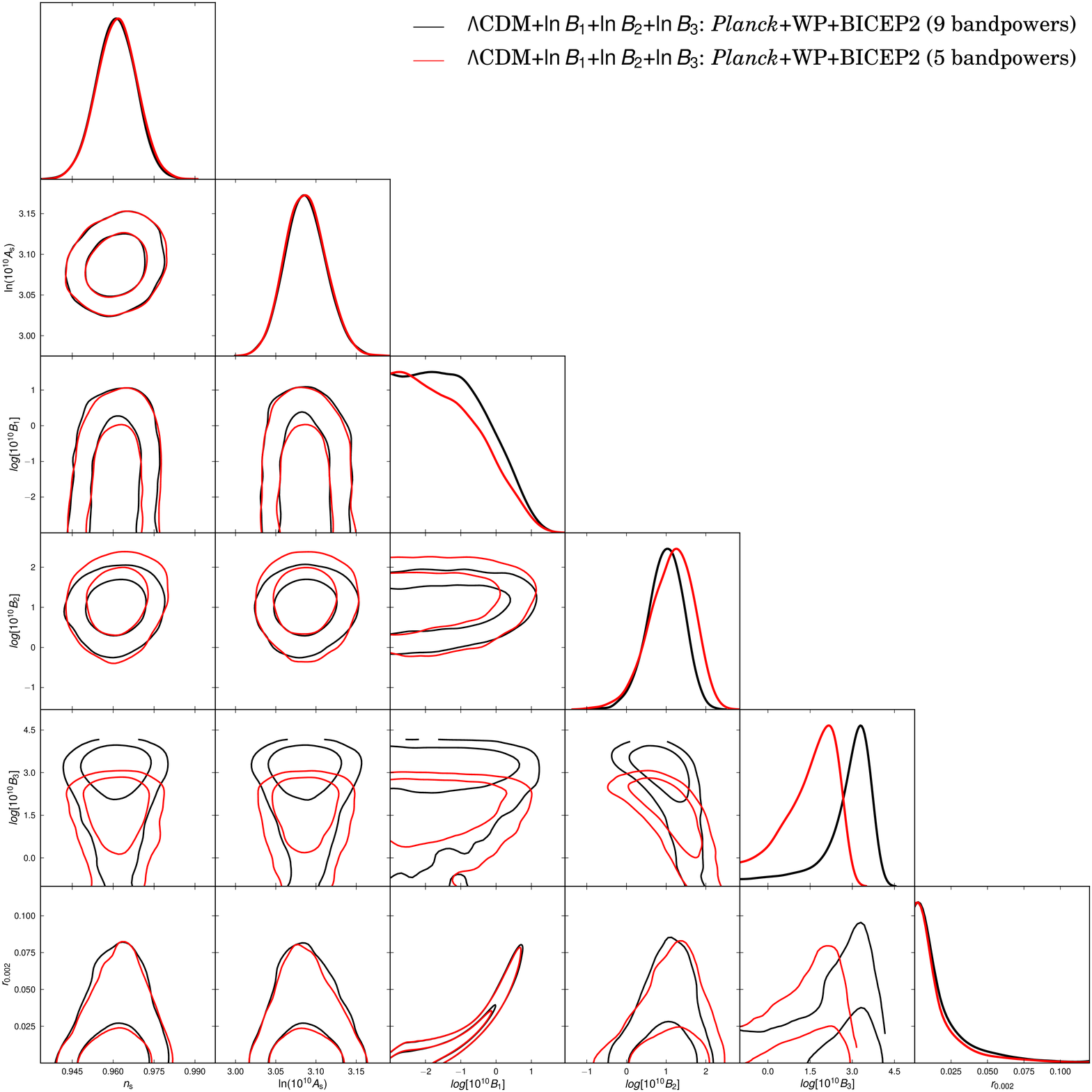}
\caption{1D/2D posterior distribution of tensor spectrum reconstruction with BICEP2.}
\label{Fig:tri_tensor_mnew}
\end{center}
\end{figure*}

\setlength\tabcolsep{1pt}
\begin{table*}[htb!]
\footnotesize
\centering
\begin{tabular}{|l|c|c|}
\hline
\multicolumn{3}{|c|}{$\Lambda$CDM+$r$+$n_t$}\\
\hline
\hline
                                            &  {\it Planck}+WP+BICEP2 (9 bandpowers)   & {\it Planck}+WP+BICEP2 (5 bandpowers) \\ 
\hline
Parameters                                  & mean $\pm$ $68\%$ C.L.            & mean $\pm$ $68\%$ C.L.\\ \hline
$100\Omega_b h^2$                           & 2.200$\pm$0.028                   & 2.204$\pm$0.028 \\
$\Omega_c h^2$                              & 0.1194$\pm$0.0026                 & 0.1195$\pm$0.0027 \\
$100\theta_{\rm MC}$                                 & 1.04129$\pm$0.00064               & 1.04127$\pm$0.00063 \\
$\tau$                                      & 0.090$\pm$0.013                   & 0.089$\pm$0.013 \\
$n_s$                                       & 0.9611$\pm$0.0073                 & 0.9615$\pm$0.0073 \\
$\ln(10^{10} A_s)$                         & 3.087$\pm$0.025                   & 3.086$\pm$0.025 \\
$r_{0.05}$                                  & $<2.00$ ($95\%$CL)                & $<2.00$ ($95\%$CL)    \\
$n_t$                                       & $1.24^{+0.51}_{-0.58}$ ($95\%$CL) & $1.20^{+0.56}_{-0.64}$ ($95\%$CL) \\
\hline
$\Omega_{m}$                                & 0.313$\pm$0.016                   & 0.313$\pm$0.016 \\
$H_0 [\mathrm{km}/\mathrm{s}/\mathrm{Mpc}]$ & 67.38$\pm$1.19                    & 67.40$\pm$1.11 \\
$r_{0.002}$                                 & $<0.061$ ($95\%$CL)               & $<0.067$ ($95\%$CL)         \\
\hline
\hline
$\chi^2_{\rm min}/2$                        & 4920.773                          & 4909.861 \\
\hline
\end{tabular}
\caption{
Mean values and $68\%$ (or $95\%$) confidence limits for primary/derived parameters in the power-law parameterization of tensor spectrum.}
\label{Tab:6rnt}
\end{table*}

\setlength\tabcolsep{1pt}
\begin{table*}[htb!]
\footnotesize
\centering
\begin{tabular}{|l|c|c|}
\hline
\multicolumn{3}{|c|}{$\Lambda$CDM+$\ln B_1$+$\ln B_2$+$\ln B_3$}\\
\hline
\hline
                                            &  {\it Planck}+WP+BICEP2 (9 bandpowers)   & {\it Planck}+WP+BICEP2 (5 bandpowers) \\ 
\hline
Parameters                                  & mean $\pm$ $68\%$ C.L.            & mean $\pm$ $68\%$ C.L.\\ \hline
$100\Omega_b h^2$                           & 2.202$\pm$0.028                   & 2.202$\pm$0.028 \\
$\Omega_c h^2$                              & 0.1194$\pm$0.0026                 & 0.1195$\pm$0.0026 \\
$100\theta_{\rm MC}$                                 & 1.04130$\pm$0.00063               & 1.04125$\pm$0.00062 \\
$\tau$                                      & 0.090$\pm$0.013                   & 0.090$\pm$0.013 \\
$n_s$                                       & 0.9610$\pm$0.0071                 & 0.9614$\pm$0.0072 \\
$\ln(10^{10} A_s)$                         & 3.087$\pm$0.025                   & 3.087$\pm$0.025 \\
$\ln(10^{10} B_1)$                         & $<0.45$ ($95\%$CL)                & $<0.39$ ($95\%$CL)    \\
$\ln(10^{10} B_2)$                         & $0.98^{+0.88}_{-0.92}$ ($95\%$CL) & $1.12^{+1.05}_{-1.02}$ ($95\%$CL) \\
$\ln(10^{10} B_3)$                         & $2.75^{+1.29}_{-2.37}$ ($95\%$CL) & $1.48^{+1.48}_{-2.00}$ ($95\%$CL) \\
\hline
$\Omega_{m}$                                & 0.313$\pm$0.016                   & 0.314$\pm$0.016 \\
$H_0 [\mathrm{km}/\mathrm{s}/\mathrm{Mpc}]$ & 67.43$\pm$1.17                    & 67.37$\pm$1.17 \\
$r_{0.002}$                                 & $<0.064$ ($95\%$CL)               & $<0.060$ ($95\%$CL)         \\
\hline
\hline
$\chi^2_{\rm min}/2$                        & 4920.558                          & 4909.799 \\
\hline
\end{tabular}
\caption{
Mean values and $68\%$ (or $95\%$) confidence limits for primary/derived parameters in the tensor spectrum cubic spline reconstruction.}
\label{Tab:BBB}
\end{table*}

\begin{figure*}[tmb]
\begin{center}
\includegraphics[scale=0.2]{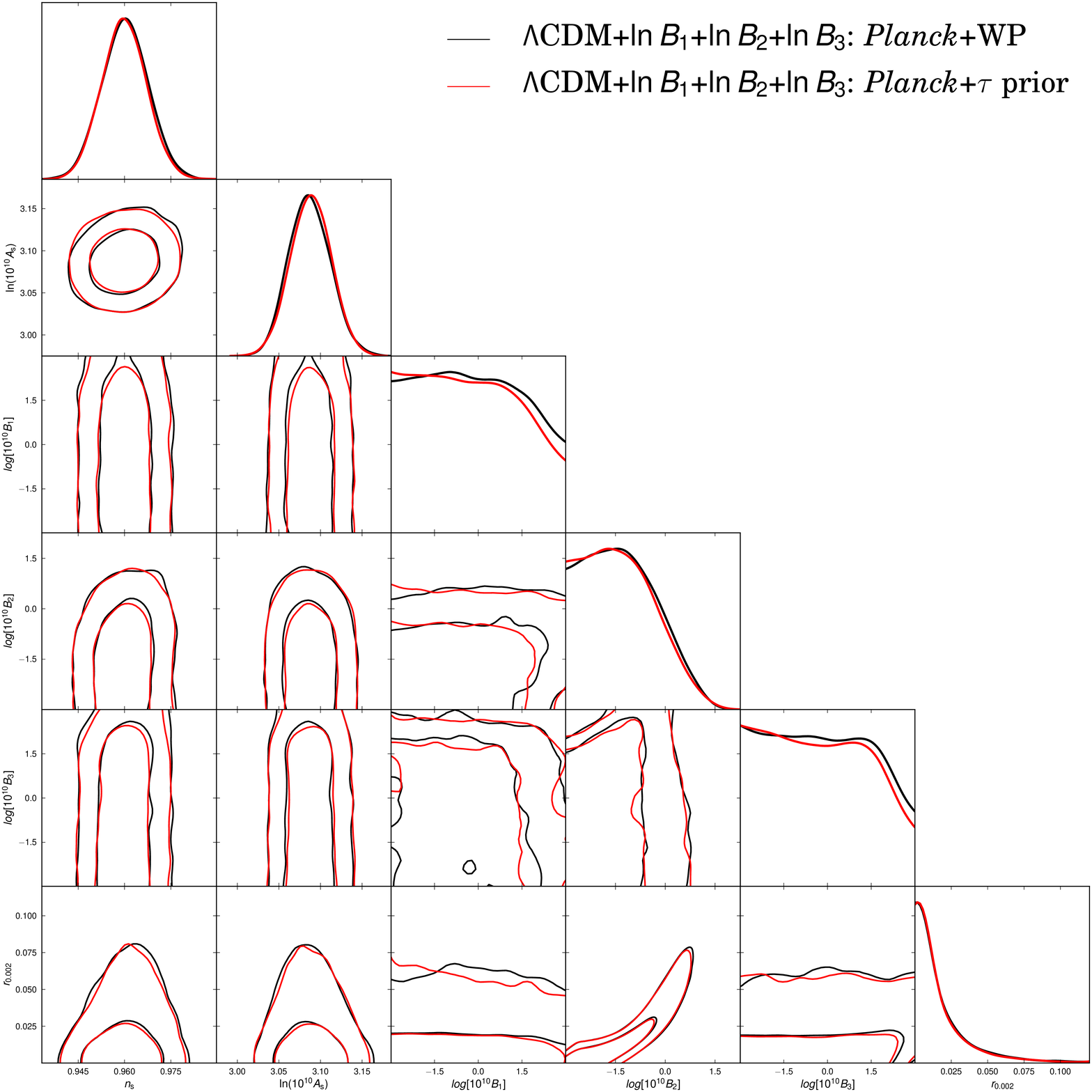}
\caption{1D/2D posterior distribution of tensor spectrum reconstruction without BICEP2.}
\label{Fig:tri_tensor_nobicep2}
\end{center}
\end{figure*}

\setlength\tabcolsep{1pt}
\begin{table*}[htb!]
\footnotesize
\centering
\begin{tabular}{|l|c|c|}
\hline
\multicolumn{3}{|c|}{$\Lambda$CDM+$\ln B_1$+$\ln B_2$+$\ln B_3$}\\
\hline
\hline
                                            &  {\it Planck}+WP   & {\it Planck}+$\tau$ Prior \\ 
\hline
Parameters                                  & mean $\pm$ $68\%$ C.L.            & mean $\pm$ $68\%$ C.L.\\ \hline
$100\Omega_b h^2$                           & 2.204$\pm$0.028                   & 2.203$\pm$0.028 \\
$\Omega_c h^2$                              & 0.1200$\pm$0.0028                 & 0.1201$\pm$0.0027 \\
$100\theta_{\rm MC}$                        & 1.04122$\pm$0.00063               & 1.04119$\pm$0.00063 \\
$\tau$                                      & 0.089$\pm$0.013                   & 0.090$\pm$0.012 \\
$n_s$                                       & 0.9603$\pm$0.0074                 & 0.9599$\pm$0.0071 \\
$\ln(10^{10} A_s)$                         & 3.087$\pm$0.025                   & 3.089$\pm$0.024 \\
$\ln(10^{10} B_1)$                  & $(-3.00,3.00)$ ($95\%$CL)                & $(-3.00,3.00)$ ($95\%$CL)    \\
$\ln(10^{10} B_2)$                         & $(-3.00,0.49)$ ($95\%$CL)         & $(-3.00,0.45)$ ($95\%$CL) \\
$\ln(10^{10} B_3)$                  & $(-3.00,3.00)$ ($95\%$CL)                & $(-3.00,3.00)$ ($95\%$CL) \\
\hline
$\Omega_{m}$                                & 0.317$\pm$0.017                   & 0.317$\pm$0.016 \\
$H_0 [\mathrm{km}/\mathrm{s}/\mathrm{Mpc}]$ & 67.20$\pm$1.22                    & 67.14$\pm$1.17 \\
$r_{0.002}$                                 & $<0.064$ ($95\%$CL)               & $<0.062$ ($95\%$CL)         \\
\hline
\hline
$\chi^2_{\rm min}/2$                        & 4902.387                          & 3895.305 \\
\hline
\end{tabular}
\caption{
Mean values and $68\%$ (or $95\%$) confidence limits for primary/derived parameters in the tensor spectrum cubic spline reconstruction.}
\label{Tab:BBB_NoBICEP2}
\end{table*}


\end{document}